\documentclass[a4paper,11pt]{article}
\pdfoutput=1 

\usepackage{epsfig, amssymb}
\usepackage{graphicx,epsfig}
\usepackage{epstopdf}
\usepackage{graphicx}
\usepackage{caption}
\usepackage{subfig}
\usepackage{color}
\usepackage{amsmath}
\usepackage{authblk}

\usepackage[T1]{fontenc} 

\title{\boldmath Quasilocal Smarr relation for an asymptotically flat spacetime \\
$\;$ \\}

\author[a]{Yein Lee \thanks{lyi126@khu.ac.kr}}
\author[b]{Matthew Richards \thanks{richam14@mcmaster.ca}}
\author[c]{Sean Stotyn \thanks{sean.stotyn@ucalgary.ca}}
\author[d]{Miok Park \thanks{miokpark@kias.re.kr}} 

\affil[a]{\it{\small{Department of Physics and Research Institute of Basic Science, Kyung Hee University, Seoul 02447, Republic of Korea}}}
\affil[b]{\it{\small{Department of Physics and Astronomy, McMaster University,1280 Main St. W., Hamilton ON L8S 4M1, Canada}}}
\affil[c]{\it{\small{Department of Physics and Astronomy, Faculty of Science, University of Calgary,2500 University Drive NW, Calgary, AB, Canada T2N 1N4}}}
\affil[d]{\it{\small{School of Physics, Korea Institute for Advanced Study,85 Hoegiro, Dongdaemun-gu, Seoul 02455, Republic of Korea}}}

\date{}

\begin{document}
\maketitle
\flushbottom

\abstract{We investigate the thermodynamics of Einstein-Maxwell(-Dilaton) theory for an asymptotically flat spacetime in a quasilocal frame. We firstly define a quasilocal thermodynamic potential via the Euclidean on-shell action and formulate a quasilocal Smarr relation from Euler’s theorem. Then we calculate quasilocal energy and surface pressure by employing Brown-York quasilocal method along with Mann-Marolf counterterm and find entropy from the quasilocal thermodynamic potential. These quasilocal variables are consistent with Tolman temperature and the entropy in a quasilocal frame turns out to be same as the Bekenstein-Hawking entropy. As a result, we found that a surface pressure term and its conjugate variable, a quasilocal area, do not participate in a quasilocal thermodynamic potential, but should present in a quasilocal Smarr relation and the quasilocal first law of black hole thermodynamics. For dyonic black hole solutions having dynamic dilaton field, non-trivial dilaton contribution should take part in the quasilocal first law but not in the quasilocal Smarr relation.}

\newpage
\tableofcontents

\section{Introduction}

In curved spacetime, conserved energy for matter fields can be obtained from the energy-momentum tensor of the Einstein equation, but this is not applied for a gravitational field. Instead, we should extract conserved energy from the metric or the Riemann tensor, but constructing the concept of gravitational energy was an arduous task in the early of general relativity due to the unification of space and time. Another difficulty was that in curved spacetime physical quantities are computed locally, whereas there is no meaningful local notion of energy density of the gravitational field to construct a conserved charge in general relativity.

Though these difficulties, the global charge for a gravitational field was first successfully obtained in 1959 by R. Arnowitt, S. Deser, and C. Misner, which is known as the ADM mass \cite{Arnowitt:1959ah,Arnowitt:1960es,Arnowitt:1960zzc,Arnowitt:1961zz}. Through this approach, spacetime is decomposed into a spatial hypersurface foliated by time so that the traditional canonical method is applied for a gravitational field at spatial infinity. The ADM result implies that a global charge is independent of a choice of a coordinate. Based on this success Komar alternatively constructed a conserved charge formula by using a Killing vector field
\begin{equation}
Q_{\xi} = \frac{c}{16 \pi G} \lim_{S_{t} \rightarrow \infty} \oint_{S_{t}} \nabla^{\alpha} \xi^{\beta} dS_{\alpha \beta}
\end{equation} 
where $c=-2$ for a timelike Killing vector field and $c=1$ for a rotational Killing vector field which give mass and angular momentum for a gravitational field respectively. Later, various methods such as AD(T) method \cite{Abbott:1981ff,Abbott:1982jh}, Brown-York's quasilocal method \cite{Brown:1992br}, and covariant phase space method \cite{Iyer:1994ys,Wald:1999wa} have been developed for calculating a global charge of a gravitational field. Nevertheless there are cases that are technically involved when applying these methods to such as, the charged rotating black hole. For that case, in 1973 \cite{Smarr:1972kt}  Larry Smarr found that a mass parameter, $M$, can be expressed by a simple algebraic relation between physical variables. By applying Euler's theorem the black hole charge $M$ is written as
\begin{equation}
M = 2 T_{\textrm{eff}} A + 2 \Omega L + \Phi Q
\label{eq:ASSmarr}
\end{equation}
where $T_{\textrm{eff}} = \frac{\tilde{\kappa}}{8 \pi}$ is an effective surface gravity, $A$ is the black hole area, $\Omega$ is an angular velocity, $L$ is an angular momentum, $\Phi$ is an electromagnetic potential, and $Q$ is a charge of the electromagnetic field. It is important to note that the mass is expressed in a bilinear form of other physical variables. The differential form of mass, $dM$, was also shown in the paper as
\begin{equation}
dM = T_{\textrm{eff}} dA + \Omega dL + \Phi dQ.
\label{eq:ASFlaw}
\end{equation}
Soon after J. Bardeen, B. Carter, and S. Hawking published the same results of (\ref{eq:ASSmarr}) and they interpreted that (\ref{eq:ASFlaw}) is analogous to the first law of thermodynamics by corresponding $4 T_{\textrm{eff}}= T = \frac{\tilde{\kappa}}{2 \pi}$ to real temperature and $S=\frac{A}{4G}$ to entropy \cite{Bardeen:1973gs}. At that time, those authors considered that $T$ and $S$ should be distinct from real temperature and entropy. Later Jacob Beskenstein argued that $S$ could be the real entropy of black holes \cite{Bekenstein:1972tm,Bekenstein:1973ur,Bekenstein:1974ax}, and Steven Hawking further corroborated this by finding that black holes radiate as thermal objects at the temperature of precisely $\frac{\tilde{\kappa}}{2 \pi}$ \cite{Hawking:1974sw,Hartle:1976tp}. 


All of this work was developed using asymptotic charges. However, it is important to describe the physical variables of a gravitational system in a quasilocal frame for several reasons. First, the quasilocal quantities could describe more realistic and detailed physical situations, such as binary stars or black hole mergers. Second, spatial infinity cannot be realized in numerical works, but rather finite domains are always required. Numerical studies of collapse also often track apparent horizons which are quasilocal in nature compared to event horizons \cite{thornburg2007event,faraoni2017foliation}. Several quasilocal formalisms have been suggested \cite{Szabados:2004vb}, and the first law of black hole thermodynamics in a quasilocal frame is studied in Brown-York's work \cite{Brown:1992br}. In such work, the subtraction background method is employed to render the gravity action finite, and the boundary energy-momentum stress tensor is constructed so as to define quasilocal quantities. They showed that the first law of thermodynamics in a finite domain for the four-dimensional Schwarzchild black hole is yielded as
\begin{equation}
d E = T_{\textrm{Tolman}} d S + P dA
\label{eq:1stlwQL}
\end{equation} 
where the Bekenstein-Hawking entropy $S$ is used, satisfying the area law. $P$ is surface pressure and $A$ is an area of a quasilocal surface with a certain radius, for example $r=R$. $E$ is a quasilocal energy and $T_{\textrm{Tolman}}$ is the Tolman temperature defined by
\begin{equation}
T_{\textrm{Tolman}} = \frac{1}{N(R)} T_{\textrm{Hawking}} = \frac{1}{N(R)} \frac{\tilde{\kappa}}{2 \pi}
\label{eq:TolmanT}
\end{equation}
where $\tilde{\kappa}$ is a surface gravity at the black hole horizon and $N(R)$ is the lapse function evaluated on the quasilocal boundary. Hereafter we denote the Tolman temperature $T_{\textrm{Tolman}}$ as $T_{\textrm{R}}$ which means the temperature measured at $r=R$ and the Hawking temperature $T_{\textrm{Hawking}}$ as $T_{\textrm{H}}$ which is measured by an infinity observer.


Historically the Smarr relation was discovered as a consequence of the scaling relations present amongst the parameters in black hole thermodynamics and can be obtained by geometric means. In this development the Smarr relation has been only associated with the Hawking temperature but not with the Tolman temperature. Thus we here aim to completely and consistently describe the Smarr relation and other thermodynamic relations by extending their notion at infinity to a quasilocal frame. In order to do so, we firstly employ the Brown-York qusilocal formalism and adopt a Mann-Marolf(MM) counterterm to construct a renormalized gravity action rather than using the subtraction method. This is because the MM-counterterm is more applicable to complicated spacetime such as the spacetime with dilaton. We also show an electric and magnetic potential form in a quasilocal frame. Next we define a quasilocal thermodynamic potential through the Euclidean action combined with the Tolman temperature to obtain entropy in a quasilocal frame. Here the quasilocal thermodynamic potential takes a different form for an electrically or magnetically charged cases. Since they impose the different boundary condition on Euclidean on-shell action the formal case yields the thermodynamic potential with the fixed gauge potential (Grand potential) but the later one yields the thermodynamic potential with the fixed charge (free energy) in \cite{Hawking:1995ap} unless introducing the boundary term for the gauge field.  From this construction we found that the surface pressure and its conjugate variable do not participate the quasilocal thermodynamic potential and the entropy in a quasilocal frame agrees with the Bekentstein-Hawking entropy. For the case of having dilaton field, we found that there is no dilaton contributions in the thermodynamic potential. Then we formulate a quasilocal Smarr relation from the Euler’s theorem. While the surface pressure and its quasilocal area take part in the relation, the dilaton field and its conjugate potential do not enter the relation since the length dimension of the dilaton field is zero. In the quasislocal first law the surface pressure terms present but dilaton terms only present for dynamical dilaton case. Thus the quasilocal thermodynamic relations can be summarized as follows
\begin{align}
&F_\textrm{R} = E -T_R S - \Phi_{\textrm{R,e}} Q_\textrm{e}\\
&E = 2T_\textrm{R} S + 2 P A + \Phi_{\textrm{R,e}} Q_\textrm{e} + \Phi_{\textrm{R,m}} Q_\textrm{m} \\
&dE = T_{\textrm{R}} dS + PdA + \Phi_{\textrm{R,e}} dQ_{\textrm{e}} + \Phi_{\textrm{R,m}} dQ_{\textrm{m}} + \Pi_{\textrm{R}} d\phi_{\textrm{R}}
\end{align}
for the four dimensional dyonic black hole with dynamic dilaton and where $\phi_R$ and $\Pi_R$ are the dilaton field and its conjugate potential in a quasilocal frame respectively, and 
\begin{align}
&F_\textrm{R} = E -T_R S - \Phi_{\textrm{R,e}} Q_\textrm{e}\\
&(n-3)E = (n-2)T_\textrm{R} S + (n-2) P A + (n-3) \Phi_{\textrm{R,e}} Q_\textrm{e}  \\
&dE = T_{\textrm{R}} dS + PdA + \Phi_{\textrm{R,e}} dQ_{\textrm{e}}  + \Pi_{\textrm{R}} d\phi_{\textrm{R}}
\end{align}
for $n$-dimensional electrically charged black hole with dynamic dilaton. For the non-dynamic dilaton case, there is no dilaton contribution in the first law. Based on these formulation, we investigate the quasilocal thermodynamics for various black hole spacetimes in Einstein-Maxwell-Dilaton theory. 
 
This paper is organized as follows. In section 2, we introduce the quasilocal formulations which are used in this paper. Firstly the Mann-Marolf counterterm and Brown-York's formalism are briefly explained along with the construction of the renormalised gravity action, and the free energy (or thermodynamic potential) in a quasilocal frame is defined by the Euclidean method to derive entropy in a quasilocal frame. Then a quasilocal Smarr relation is constructed from the Eulerian theorem for an electrically charged black holes in $n$ dimensional spacetime. In section 3 and 4, we check thses quasilocal thermodynamic relations with various black hole solutions from the Einstein-Maxwell(-Dilaton) theory. Lastly, we summarize our results and discuss the future works. 


\section{The quasilocal formulation}

In this section we provide the quasilocal thermodynamic relations. Through this paper we consider $n$-dimensional spacetime $(\mathcal{M},g)$ and denote its index $\mu=0, \cdots ,n-1$. The timelike boundary $(\partial \mathcal{M},h)$ is a timelike hypersurface defined by a spacelike normal vector $n^{\mu}$ and we denote its index $a=0,2, \cdots, n-1$. Finally, the spatial boundary of a constant time slice $(B,\sigma)$ is a spacelike hypersurface which is normal to both of the spacelike normal vector $n^{\mu}$ and a timelike normal vector $u^{a}$ and we denote its index $A=2, \cdots, n-1$. Physically, $B$ is the geometry of the quasilocal boundary and $\partial \mathcal{M}$ is the evolution of $B$ through time. 

\subsection{Mann-Marolf counterterm}

It is known that a gravity action diverges for non-compact spacetime as $r$ goes to infinity, while compact spacetime does not have such divergence. For an asymptotically flat spacetime, this divergence is mainly due to considering a surface term which is known as the Gibbons-Hawking (GH) term. As a remedy, a non-dynamical term is introduced in the action as to remove the divergence. Here the non-dynamical term should not alter the equation of motion but render the total gravity action finite
\begin{equation}
I_{\textrm{remormailzed}} = I_{\textrm{EH}} + I_{\textrm{GH}} + I_{\textrm{non-dynamical}}.
\end{equation}

To generate this non-dynamical term, the reference background approach was suggested in \cite{Hawking:1995fd}. However, if the dimension of spacetime is higher than three, the existence or uniqueness of such embeddings of a hypersurface $(\partial \mathcal{M}, h)$ into a proper reference frame $(\mathcal{M}_{\textrm{Ref}},g_{\textrm{Ref}})$ is not clear. On the other hand, as the interest in AdS/CFT theory has increased, the algorithm to generate counterterms as a non-dynamical term have been well constructed for AdS spacetime \cite{Balasubramanian:1999re,deBoer:1999tgo,Bianchi:2001kw,Papadimitriou:2016yit}. The same algorithm for generating counterterms, unfortunately, is not applicable to an asymptotically flat spacetime. However, the Mann-Marolf(MM) counterterm method provides one way to generate counterterms for an asymptotically flat spacetime and is described as follows.

The gravity action is written as
\begin{equation}
I = \frac{1}{\kappa^{2}} \int_{\mathcal{M}} \sqrt{-g} R + \frac{2}{\kappa^{2}} \int_{\partial \mathcal{M}} \sqrt{-h} (K - \hat{K})
\label{eq:MMactn}
\end{equation}
where $\hat{K}$ is a solution of
\begin{equation}
\mathcal{R}_{ab} = \hat{K}_{ab} \hat{K} - \hat{K}_{a}^{\; c} \hat{K}_{cb}
\label{eq:MMrltn}
\end{equation}
where $\mathcal{R}_{ab}$ is the Ricci tensor on the boundary $\partial \mathcal{M}$. This counterterm is known to be local and covariant. The relation (\ref{eq:MMrltn}) is motivated from the Gauss-Codazzi equation for a timelike hypersurface which is described as follows 
\begin{equation}
\mathcal{R}_{acbd} = R^{\textrm{Ref}}_{acbd} + K_{ab}K_{cd} - K_{cb}K_{ad}
\label{eq:GC}
\end{equation}
where $\mathcal{R}_{acbd}$ is the Riemann tensor on $\partial \mathcal{M}$ compatible with the induced metric $h$ and $R^{\textrm{Ref}}_{acbd}$ is the Riemann tensor of $(\mathcal{M}_{\textrm{Ref}},g_{\textrm{Ref}})$ pulled back to $\partial \mathcal{M}$, and is obtained by taking advantage of a reference background approach on a hypersurface near the infinity. Taking the Minkowski spacetime as a reference frame then $R^{\textrm{Ref}}_{acbd}$ becomes zero and contracting (\ref{eq:GC}) with $g^{cd}$ it yields (\ref{eq:MMrltn}). 

Considering a general form of the metric in $n$ dimensional spacetime 
\begin{align}
ds^{2} &= N^{2} dr^{2} + h_{ab} dx^{a} dx^{b} \nonumber\\
&= \bigg(1+ \frac{\alpha}{r^{n-3}} \bigg)^{2} dr^{2} +  \bigg( h^{(0)}_{ab} + \frac{1}{r^{n-3}}h^{(1)}_{ab} + \frac{1}{r^{n-2}} h^{(2)}_{ab} + \cdots \bigg)dx^{a} dx^{b}
\label{eq:MMansatz}
\end{align}
where
\begin{align}
&h_{ab} dx^{a} dx^{b} =\nonumber\\
 &- \bigg( 1 + \frac{\gamma^{(1)}}{r^{n-3}} + \frac{\gamma^{(2)}}{r^{n-2}} + \cdots \bigg) dt^{2} + r^{2} \bigg(\mu^{(0)}_{AB} + \frac{1}{r^{n-3}} \mu^{(1)}_{AB} + \frac{1}{r^{n-2}} \mu^{(2)}_{AB} + \cdots \bigg) d \eta^{A} d \eta^{B}.
\end{align}
Choosing $\partial \mathcal{M}$ to be cylindrical $\Omega^{\mathrm{cyl}}$ such as 
\begin{equation}
\Omega^{\mathrm{cyl}} = r + O(r^{0}),
\end{equation}
the solution of (\ref{eq:MMrltn}) is calculated each for four dimensions and higher than four dimensions 
\begin{align}
(n=4) \; \; \hat{K}_{ab} &= r \mu^{(0)}_{ab} + \frac{1}{2} \mu^{(1)}_{ab} + D_{a}D_{b} \alpha + \cdots \label{eq:MMn4} \\
(n>4) \; \; \hat{K}_{ab} &= r \mu^{(0)}_{ab} + \frac{1}{(n-4) r^{n-5}} \bigg[ (n-4)\mu^{(1)}_{ab} + \alpha \mu^{(0)}_{ab} + \gamma^{(1)} \mu^{(0)}_{ab} +D_{a} D_{b} \alpha \bigg]  + \cdots
\label{eq:MMnb4}
\end{align}
where the higher sub-leading terms of order of $r$ are omitted here, but up to the sub-sub-leading term of order of $r$ is computed in \cite{Park:2012bv}. In the asymptotically flat case, the leading term of the counterterm removes divergence of the gravity action and sub-leading terms give corrections to the finite parts of the action. In this paper we consider up to the first sub-leading term of the MM-counterterm solutions, as displayed in (\ref{eq:MMn4}) and (\ref{eq:MMnb4}). Even in the presence of matter fields, the MM-counterterm (\ref{eq:MMn4}) and (\ref{eq:MMnb4}) can be still utilized, as long as the matter fields do not generate any divergent behaviour in the action as $r$ goes to infinity and a metric form is taken into (\ref{eq:MMansatz}).


\subsection{BY method and electic/magnetic charge potential}

The Hawking temperature is defined relative to an observer located where a timelike Killing vector field has unit norm $\sqrt{-\xi^{\mu} \xi_{\mu}} = 1$, which indicates the observer is located at infinity in an asymptotically flat spacetime. As a consequence, any stationary observer at finite radius will measure a redshifted temperature known as Tolman temperature (\ref{eq:TolmanT}). Associated with this local temperature, quasilocal thermodynamic quantities should be defined accordingly. 

One of the quasilocal formalism is constructed by Brown and York in \cite{Brown:1992br}. Here we employ their method. In order to construct the renormalized action, the MM-counterterm is taken as follows
\begin{equation}
I_{\textrm{remormailzed}} = I_{\textrm{EH}} + I_{\textrm{GH}} + I_{\textrm{MM-counterterm}}
\label{eq:reSmm}
\end{equation}
and this produces the Brown-York's boundary stress energy-momentum tensor, which is defined as
\begin{equation}
\tau^{ab} = \frac{2}{\sqrt{-h}} \frac{\delta I_{cl}}{\delta h_{ab}} = \frac{2}{\kappa^{2}}(\pi^{ab} - \hat{\pi}^{ab}) 
\label{eq:EMtns}
\end{equation}
where $I_{cl}$ is the on-shell action of $I_{\textrm{remormailzed}}$, $\pi_{ab} = K_{ab} - K h_{ab}$, and $\hat{\pi}_{ab} = \hat{K}_{ab} - \hat{K} h_{ab}$. This tensor yields a quasilocal energy density $\epsilon$, proper momentum surface density $j_{A}$ and boundary stress $s^{AB}$ as follows
\begin{align}
\varepsilon \equiv u_{a} u_{b} \tau^{ab}, \qquad j_{A} \equiv - \sigma_{Aa}u_{b} \tau^{ab}, \qquad s^{AB} \equiv \sigma^{A}_{a} \sigma^{B}_{b} \tau^{ab}
\label{eq:QuasilocalQ}
\end{align}
where $u_{a}$ is a timelike normal vector field, $\sigma_{AB}$ is an induced metric on the hypersurface $B$. In $n$-dimensional spacetime, surface pressure is defined by
\begin{equation}
P = - \frac{1}{(n-2)} \tau_{ab} \sigma^{ab}
\label{eq:QuasilocalP}
\end{equation}
where $\sigma^{ab}$ is the pull-back of $\sigma^{AB}$. The quasilocal energy contained within $B$ is obtained by surface integration 
\begin{equation}
E = \int_{B} d^{n-2} x \sqrt{\sigma} \varepsilon,
\label{eq:QLE}
\end{equation}
and the conserved quantity along the Killing vector field, $\xi$, is defined by
\begin{equation}
Q_{\xi} = \int_{B} d^{n-2} x \sqrt{\sigma} (\varepsilon u^{i} + j^{i}) \xi_{i}.
\end{equation}


In the presence of matter fields such as a non-abelian gauge field ${\mathcal{A}}_{\mu}$, the total electric charge $Q_\textrm{e}$ which is confined inside of an event horizon is defined as
\begin{equation}
Q_{\textrm{e}} = \int *{\mathcal{F}}
\label{eq:totalelecQ}
\end{equation}
where ${\mathcal{F}}$ is a field strength, which is 
\begin{equation}
{\mathcal{F}}_{\mu \nu} = \partial_{\mu} {\mathcal{A}}_{\nu} - \partial_{\nu} {\mathcal{A}}_{\mu}, \qquad *{\mathcal{F}} = \frac{1}{2} \epsilon_{\mu \nu \alpha \beta} {\mathcal{F}}^{\alpha \beta} dx^{\mu} \wedge dx^{\nu}
\end{equation} 
An electric potential measured by an observer with a four velocity $u^{\mu}$ is given by \cite{Creighton:1996st}
\begin{equation}
\Phi_{\textrm{e}} (r) = {\mathcal{A}}_{\mu}(r) u^{\mu}(r) \bigg|^{r_{h}}_{r}.
\label{eq:QLEP}
\end{equation}
A magnetic charge in four dimensions is defined as 
\begin{equation}
Q_{\textrm{m}} = \int {\mathcal{F}},
\label{eq:totalmagenticQ}
\end{equation}
and due to the electro-magnetic duality we can construct ${\mathcal{A}}^{*}_{\mu}$ as follows
\begin{align}
*{\mathcal{F}}_{\mu \nu} = \partial_{\mu} {\mathcal{A}}^{*}_{\nu} - \partial_{\nu} {\mathcal{A}}^{*}_{\mu}\qquad 
\label{eq:emdual}
\end{align}
and a magnetic potential measured by an observer with a four velocity $u^{\mu}$ can be defined as
\begin{equation}
\Phi_{\textrm{m}}(r) = {\mathcal{A}}^{*}_{\mu}(r) u^{\mu}(r) \bigg|^{r_{h}}_{r}. 
\label{eq:QLMP}
\end{equation}

\subsection{Quasilocal thermodynamic potential}

Here we aim to find black hole entropy in a quasilocal frame. In order to do so, we define a quasilocal thermodynamic pontential via the Euclidean method. 

Black hole thermodynamics can be understood by path integral at finite temperature in Euclidean space, which is generated by doing a Wick rotation $\tau \rightarrow i t$ from the Lorentzian one. Then the partition function is written as
\begin{equation}
Z = \int D[g] D[\Psi] e^{-I_{E}[g,\Psi]/\hbar}
\end{equation}
where $g$ is a fluctuation of the metric and $\Psi$ are matter fields. Taking a saddle point approximation the partition function is approximated to 
\begin{equation}
Z \sim e^{-I_{E}[g,\Psi]/\hbar}.
\end{equation}
In the calculation of the Euclidean action $I_E$, when we consider the boundary of spacetime to the infinity, the radial coordinate $r$ is integrated from the black hole horizon to infinity. The Euclidean time integration should give the periodicity $\beta$ at the black hole horizon to avoid a conical singularity. Then it yields an inverse of Hawking temperature, $\int d \tau = \beta = \frac{1}{T_{\textrm{H}}}$, which is consistent with the surface gravity having unit norm of the timelike Killing vector field. That is, the Euclidean time periodicity $\beta$ is imposed at the horizon relative to the infinity observer. Then these define the free energy (or thermodynamic potential) as follows
\begin{align}
F = - T_{\textrm{H}} \log Z \sim T_{\textrm{H}} I_{E,(r_{h},\infty)} = M - T_{\textrm{H}} S (- \Phi_{\textrm{e}} Q_{\textrm{e}}).
\end{align}
where $M$ is the conserved charge and $\Phi_{\textrm{e}}$ is measured between the horizon and infinity. 

Now we extend this formula to a quasilocal frame by considering the boundary of spacetime at the finite domain $r=R$. The radial coordinate $r$ is integrated from the horizon to the finite distance of $r=R$ and then the time periodicity should be relative to the observer at $r=R$, which is described by $\int d \tau' = N(R)\int d\tau= N(R)\beta = \frac{1}{T_{\textrm{R}}}$. Thus the Hawking temperature is replaced by the Tolman temperature. By doing so we define quasilocal free energy as follows
\begin{align}
F_{R} \equiv - T_{\textrm{R}} \log Z \sim T_{\textrm{R}} I_{E, (r_{h}, R)} = E - T_{\textrm{R}} S (- \Phi_{\textrm{R,e}} Q_{\textrm{e}})
\label{eq:DnfreeEQL}
\end{align}
where $E$ is quasilocal energy and the electric potential, $\Phi_{\textrm{R,e}}$, is measured to the finite distance $r=R$. By obtaining this quasilocal free energy from the Euclidean on-shell action value we can derive entropy in a quasilocal frame, which turns out to be same as the Bekenstein-Hawking entropy from our examples in next sections. 

\subsection{Quasilocal Smarr relation by Euler's Theorem}

In \cite{Smarr:1972kt}, Euler's theorem on homogeneous functions is used to obtain the Smarr relation. Euler's theorem states that if a function $f(x, y,z)$ obeys the scaling relation
\begin{equation}
	f(\alpha^p x, \alpha^q y, \alpha^k z) = \alpha^r f(x,y,z),
\end{equation}
it satisfies
\begin{equation}
r f(x,y,z) = p \left(\frac{\partial f}{\partial x} \right) x + q \left(\frac{\partial f}{\partial y} \right) y + k \left(\frac{\partial f}{\partial z} \right) z.
\end{equation}
Applying this theorem into black hole systems for Einstein-Maxwell theory in four dimensions, a black hole charge $M$ can be considered as a homogeneous function of entropy $S$ and the Maxwell charge $Q$. These variables have the following scaling properties
\begin{equation}
	M \propto [L], \; \; \; S \propto [L]^{2}, \; \; \; Q \propto [L]
\end{equation}
where $[L]$ is length dimension. Then the Euler's theorem yields
\begin{align}	
M =& 2 \bigg(\frac{\partial M}{\partial S} \bigg) S + \bigg(\frac{\partial M}{\partial Q} \bigg) Q = 2 T_{\textrm{H}} S + \Phi Q.
\label{eq:SmarrE}
\end{align}
Note that here we employ the natural unit, which is $c=\hbar=1$ and additionally impose the gravitational constant $G=1$, which leads the action to be dimensionful, $I \propto [L]^{2}$. 

Let us extend this argument into a quasilocal frame. In this case a family of timelike hypersurfaces exists for $r=constant$ and one specific hypersurface can be chosen, for example $r=R$. Thus the quasilocal energy $E$ becomes a homogenous function of entropy $S$ and the Maxwell charge $Q$, but also a quasilocal area $A$ whose radius is $R$. Their scaling properties are found as
\begin{equation}
	E \propto [L], \; \; \; S \propto [L]^{2}, \; \; \; Q \propto [L], \; \; \; A \propto [L]^{2}.
\end{equation}
So the Euler's theorem yields the following relation 
\begin{align}	
E =& 2 \bigg(\frac{\partial E}{\partial S} \bigg) S + \bigg(\frac{\partial E}{\partial Q} \bigg) Q + 2 \bigg(\frac{\partial E}{\partial A} \bigg) A = 2 T_{\textrm{R}} S + \Phi_{\textrm{R}} Q + 2 P A .
\label{eq:quasiSmarr}
\end{align}
where a new term $PA$ arises. When $r$ goes to infinity the surface pressure $P$ vanishes and the quasilocal Smarr relation is restored to (\ref{eq:SmarrE}). The electric potential (\ref{eq:QLEP}) should be measured by an observer placed at $r=R$ to be consistent with the Tolman temperature (\ref{eq:TolmanT}). When considering Einstein-Maxwell-Diltaton theory we should investigate the participation of a dilaton field, $\phi(r)$, to the quasilocal Smarr relation. The conjugate variable of the dilaton field is defined by \cite{Creighton:1995au}
\begin{equation}
\Pi(r)= \int d^{2}x \sqrt{\sigma} n_{\mu} \partial_{\nu} \phi g^{\mu \nu}, \qquad \Pi_{\textrm{R}} = \Pi(R)-\Pi(\infty) \label{eq:PiR}
\end{equation}
where the dilaton potential $\Pi_{\textrm{R}}$ is calibrated to be zero at infinity. However since the length dimension of the dilaton field is $\phi \propto [L]^{0}$ it does not contrube to the quasilocal Smarr relation. 

If we expand this argument into $n$-dimensional spacetime, each variable has a length dimension as
\begin{equation}
E \propto [L]^{n-3}, \; \; \; S \propto [L]^{n-2}, \; \; \; A \propto [L]^{n-2}, \; \; \; Q \propto [L]^{n-3},  \; \; \; \phi \propto [L]^{0}
\end{equation}
and Euler's theorem yields the relation as follows
\begin{align}
(n-3) E =& (n-2) \left(\frac{\partial E}{\partial S} \right) S + (n-3)  \left(\frac{\partial E}{\partial Q} \right) Q + (n-2) {\left(\frac{\partial E}{\partial A} \right) A} \\
=& (n-2) T_{\textrm{R}} S + (n-3) \Phi_{\textrm{R}} Q_ + (n-2) PA 
\label{eq:QLSmarr}
\end{align}
which is the $n$-dimensional quasilocal Smarr relation for Einstein-Maxwell(-Dilaton) theory. The thermodynamic first law associated with this Smarr relation in a quasilocal frame is written as
\begin{equation}
d E = T_{\textrm{R}} dS + \Phi_{\textrm{R}} d Q + P d A + \Pi_{\textrm{R}} d\phi_{\textrm{R}}
\label{eq:QLfl}
\end{equation}
where $\phi_{\textrm{R}} = \phi(R)$. Even though the dilaton field does not participate in the quasilocal Smarr relation it makes an effect to the quasilocal first law \cite{Creighton:1995au}. 

These results are compared with the thermodynamic properties for Einstein-Maxwell(-Dilaton) theory at infinity which are described by
\begin{align}
&(n-3) M = (n-2) T_{\textrm{H}} S + (n-3) \Phi Q, \label{eq:SMInf}\\
&dM = T_{\textrm{H}} d S + \Phi d Q \label{eq:flInf}
\end{align}
where $\Phi$ is measured from the horizon to infinity. We examine these quasilocal Smarr relation (\ref{eq:QLSmarr}) and the first law (\ref{eq:QLfl}) with black hole solutions for Einstein and Einstein-Maxwell theory in $n$-dimensional spacetime in section 3 and for Einstein-Maxwell-Dilaton theory in four dimensional spacetime in section 4. 

\section{Einstein-Maxwell Theory}

In this section, we consider several black hole solutions for Einstein(-Maxwell) theory in an asymptotically flat spacetime and check if our quasilocal formulation is satisfied. Firstly we start with a four dimensional Schwarzschild black hole solution, and find its quasilocal first law along with the quasilocal Smarr relation. We also explore the Reissner-Nordstr{\"o}m black hole and its higher dimensional generalization. 

\subsection{Schwarzschild Black Holes}
	Let us write the Schwarzschild black holes spacetime ($\mathcal{M},g$) in the form of 
	\begin{equation}
	ds^{2} = - N(r)^{2} dt^{2} + f(r)^{2} dr^{2}  + r^{2} d \Omega_{2} 
	\label{eq:Fourmetric}
	\end{equation}
	where 
	\begin{equation}
	N(r) = \frac{1}{f(r)} = \sqrt{1 - \frac{2M}{r}}.
	\end{equation}
Assuming an observer on a hypersurface ($\mathcal{\partial M}, h$) at $r=R$, the temperature measured by this observer is given by (\ref{eq:TolmanT})
	\begin{equation}
	T_{\textrm{R}} = \frac{1}{N(R)} \frac{1}{4 \pi r_{h}}.
	\label{eq:SchTolman}
	\end{equation}
The quasilocal energy density and surface pressure are obtained as follows
	\begin{align}
	&\varepsilon=u_{i}u_{j} \tau^{ij} =  \frac{4}{\kappa^{2}}  \bigg( \frac{1}{r} - \frac{1}{r f(r)} \bigg)\bigg|_{r=R} = \frac{4}{\kappa^{2}} \bigg(\frac{1}{R} - \frac{1}{R f(R)} \bigg), 
	\label{eq:QLed} \\
	&P  = - \frac{1}{2} \sigma_{ab} s^{ab} = -\frac{2}{\kappa^{2}} \bigg(\frac{N'(r)}{f(r)N(r)} + \frac{1}{r f(r)} - \frac{1}{r} \bigg) \bigg|_{r=R} = -\frac{2}{\kappa^{2} R} \bigg(\frac{1}{N(R)} \bigg(1 - \frac{r_{h}}{2R} \bigg) - 1 \bigg), 
	\label{eq:QLsp}
	\end{align}
	where  
	\begin{equation}
	N(R) = \frac{1}{f(R)} =  \sqrt{1 - \frac{r_{h}}{R} }
	\end{equation}
and $N(r_{h}) = 0$ is used. The total quasilocal energy is obtained by integrating the quasilocal energy density (\ref{eq:QLed}) over the hypersurface at constant time ($B, \sigma$) and is yielded as
	\begin{equation}
	E(r_{h}, R) = \int_{B} d^{2} x \sqrt{\sigma} \varepsilon = \frac{16 \pi R}{\kappa^{2}} \bigg( 1 - N(R) \bigg)
	\end{equation}
which approaches the conserved charge $M$ when taking $R \rightarrow \infty$ and $G=1$. 

To obtain the entropy in a quasilocal frame, we calculate the quasilocal free energy (\ref{eq:DnfreeEQL}) from the Euclidean gravity action 
\begin{align}
F_{R} = T_{R} I_{E} = \frac{8 \pi}{\kappa^{2} N(R)} \bigg[ - \frac{r_{h}}{2} - 2 R \bigg( 1 - \frac{r_{h}}{R} - \sqrt{1 - \frac{r_{h}}{R}} \bigg) \bigg],
\end{align}
and then the entropy is yielded as
\begin{equation}
S = -\frac{F_{R} - E}{T_{R}} = \frac{16 \pi^{2} r_{h}^{2}}{\kappa^{2}}
\end{equation}
which agrees with the Bekenstein-Hawking entropy. Then we find that these quasilocal variables agree with the quasilocal Smarr relation in (\ref{eq:quasiSmarr}) with $Q=0$
\begin{equation*}
E = 2 T_{\textrm{R}} S + 2 P A.
\end{equation*} 

Now let us make a variation of $E$ with respect to $S$ and $A$ and then it simply reproduces the first law of black hole thermodynamics 
	 \begin{equation}
	 dE = T_{\textrm{R}}  d S + P d A
	 \end{equation}
where $A = 4 \pi R^{2}$.


\subsection{Reissner-Nordstr{\"o}m Black Holes}

In the presence of abelian gauge field, the Einstein equation with the metric ansatz (\ref{eq:Fourmetric}) yields the metric function as
\begin{equation}
	N(r) = \frac{1}{f(r)} = \sqrt{1 - \frac{2M}{r} + \frac{q_{i}^{2}}{r^{2}} }
\label{eq:RNmetric}
\end{equation}
where the index $i$ indicates an electrically charged case with the electric charge parameter $q_1=q$, a magnetically charged case with the magnetic charge parameter $q_2=p$ or presence of both charges. The gauge field solutions become as
\begin{equation}
{\mathcal{A}}  = \pm \frac{2}{\kappa} \bigg(\frac{q}{r} - \frac{q}{r_{h}} \bigg)dt, \qquad {\mathcal{A}}  = \pm \frac{2}{\kappa} p (1-\cos \theta) d \phi.
\end{equation}
Here the regularity condition is imposed so as to $A_{t} =0$ at black hole horizon and another gauge patch is glued for $A_{\phi}$ to be regular on the $\theta$-axis. The total electric or magnetic charge are computed from (\ref{eq:totalelecQ}) and (\ref{eq:totalmagenticQ}) respectively 
\begin{equation}
Q_E = \pm \frac{8 \pi}{\kappa} q, \qquad Q_M = \pm \frac{8 \pi}{\kappa} p. 
\end{equation}
The Tolman temperature is calculated as
	\begin{equation}
	T_{\textrm{R}}  = \frac{1}{N(R)} \frac{1}{4 \pi r_{h}} \bigg(1- \frac{q_{i}^{2}}{r_{h}^{2}} \bigg).
	\end{equation}
To compute the boundary stress energy-momentum tensor (\ref{eq:EMtns}), we employ the MM-counterterm up to the sub-leading order of $r$. Then from the renormalized action we calculate the quasilocal energy (\ref{eq:QLE}), surface pressure (\ref{eq:QuasilocalP}), and electric (\ref{eq:QLEP}) or magnetic potential (\ref{eq:QLMP}) as follows
	\begin{align}
	&E(r_{h}, R, q_{j}) = \int_{B} d^{2} x \sqrt{\sigma} \varepsilon = \frac{16 \pi R}{\kappa^{2}} \bigg( 1  - N(R) \bigg)	\label{eq:RNQLE}, \\
	&P  =- \frac{1}{2} \sigma_{ab} s^{ab} = - \frac{2}{\kappa^{2} R} \bigg(\frac{1}{N(R)} \bigg(1 - \frac{r_{h}}{2R} - \frac{q_{j}^{2}}{2 R r_{h}} \bigg) - 1 \bigg), \\
	&\Phi_{\textrm{R},i} = \pm \frac{1}{N(r)} \frac{2}{\kappa} \bigg( \frac{q_{i}}{r} - \frac{q_{i}}{r_{h}} \bigg) \bigg|^{r=r_{h}}_{r=R} = \pm \frac{1}{N(R)} \frac{2}{\kappa}  \bigg( \frac{q_{i}}{r_{h}} -  \frac{q_{i}}{R}  \bigg)
	\end{align}
	where  
	\begin{equation}
	N(R) = \frac{1}{f(R)} =  \sqrt{1 - \frac{r_{h}}{R} - \frac{q_{i}^{2}}{r_{h}R} + \frac{q_{i}^{2}}{R^{2}} }
	\end{equation}
and $N(r_{h}) = 0$ is used. The quasilocal thermodynamic potential (\ref{eq:DnfreeEQL}) obtained from the Euclidean gravity action becomes
\begin{align}
F_{R} =& T_{R} I_{E} \nonumber\\
=& \frac{8 \pi}{\kappa^{2}} \bigg[ \frac{1}{N(R)} \bigg(\frac{q_{j}^{2}}{R} - \frac{q_{j}^{2}}{r_{h}} \bigg) - \frac{1}{N(R)} \bigg(M - \frac{q_{j}^{2}}{R}\bigg) - 2 R \bigg(N(R) -1 \bigg) \bigg]\\
\equiv &E - T_{\textrm{R}}S - \Phi_{\textrm{R},i} Q_{i}
\end{align}
and the entropy is computed as 
\begin{align}
S= - \frac{F_{R} - (E-\Phi_{\textrm{R}, i} Q_{i})}{T_{R}} = \frac{16 \pi^{2} r_{h}^{2}}{\kappa^{2}}
\end{align}
which agrees with the Bekenstein-Hawking entropy. From these quasilocal variables we find that our Smarr relation (\ref{eq:quasiSmarr}) is satisfied and also confirm that the variation of quasilocal energy (\ref{eq:RNQLE}) with respect to $S$, $Q$, and $A$ retains the first law
	\begin{align}
	&E = 2 T_R S +  \Phi_{\textrm{R},i} Q_{i} + 2 P A, \\
	&dE = T_{\textrm{R}} d S + P d A + \Phi_{\textrm{R},i} d Q_{i}.
	\end{align}

\subsection{Higher dimensional electrically charged Black Holes}

Let us consider a higher dimensional charged black hole with the following metric ansatz
\begin{equation}
ds^{2} = - N(r)^{2} dt^{2} +  f(r)^{2} dr^{2} + r^{2} d \Omega_{n-2}.
\label{eq:metricHD}
\end{equation}
By solving the Einstein equation, we obtain the metric and the electric potential
\begin{align}
N(r) =& \frac{1}{f(r)} = \sqrt{1-\frac{\mu}{r^{n-3}} + \frac{q^{2}}{r^{2(n-3)}} }, \\
{\mathcal{A}}(r) =& \pm \frac{1}{\kappa} \sqrt{\frac{2(n-2)}{n-3}} \bigg( \frac{q}{r^{n-3}}-\frac{q}{r_{h}^{n-3}}  \bigg) dt
\end{align}
where the regularity condition is also imposed at the black hole horizon for ${\mathcal{A}}_t$, and the total electric charge is computed as
\begin{equation}
Q = \pm \frac{1}{\kappa}\sqrt{2(n-2)(n-3)} q  \omega_{n-2}.
\end{equation}
From the metric, Tolman temperature at $r=R$ is easily read off
\begin{equation}
T_{\textrm{R}} = \frac{1}{N(R)} \frac{(n-3)}{4 \pi r_{h}} \frac{r_{h}^{2(n-3)} - q^{2}}{r_{h}^{2(n-3)}}.
\end{equation}
To construct the renormalized gravity action, we take the MM-counterterm solutions (\ref{eq:MMnb4}) with the metric (\ref{eq:metricHD}) which is yielded as
\begin{align}
\hat{K}_{ij} = r \mu^{(0)}_{ij}
\label{eq:RNcc}
\end{align}
where the first sub-leading term vanishes since $\alpha = - \gamma^{(1)}$ in this case. Then the quasilocal energy and surface pressure are computed as
\begin{align}
E =& \int_{B} d^{n-2} x \sqrt{\sigma} \epsilon = \frac{2(n-2)}{\kappa^{2}} \omega_{n-2} R^{n-3} \bigg[1-N(R) \bigg], \\
P =&- \frac{1}{(n-2)} \tau_{ij} \sigma^{ij} \bigg|_{r=R} = - \frac{2}{\kappa^{2}} \frac{(n-3)}{R} \bigg[ \frac{1}{N(R)} \bigg(1-\frac{\mu}{2 R^{n-3}} \bigg) -1 \bigg], 
\end{align}
and when taking $R$ to infinity we obtain $E \approx \frac{(n-2) \omega_{n-2}}{\kappa^{2}} \mu$ which agrees with one in \cite{Myers:1986un}. The quasilocal electric potential at $r=R$ is written as
\begin{align}
\Phi_{\textrm{R}} =& \pm \frac{1}{N(R)} \frac{1}{\kappa} \sqrt{\frac{2(n-2)}{n-3}} \bigg( \frac{q}{r_{h}^{n-3}} - \frac{q}{R^{n-3}} \bigg).
\end{align}
where
\begin{align}
N(R) = \sqrt{1-\frac{\mu}{R^{n-3}} + \frac{q^{2}}{R^{2(n-3)}}} .
\end{align}
The quasilocal thermodynamic potential (\ref{eq:DnfreeEQL}) is calculated as
\begin{align}
F_{R} =& T_{\textrm{R}} I_{\textrm{E}} \\
=& \frac{2 \omega_{n-2}}{\kappa^{2}} \bigg[ \frac{1}{N(R)} \bigg(\frac{(n-2)  q^2}{R^{n-3}}  - \frac{(n-3)r_h^{n-3}}{2}  - \frac{(n-1)q^2}{2 r_h^{n-3}}\bigg) + (n-2) R^{n-3} \bigg(1 - N(R) \bigg)\bigg] \\
\equiv & E - T_{\textrm{R}} S - \Phi_{\textrm{R}} Q 
\label{eq:freeEnD}
\end{align}
and so the entropy is derived as 
\begin{align}
S = - \frac{F_{R} - (E - \Phi_{R}Q)}{T_{R}} = \frac{4 \pi }{\kappa^{2}} \omega_{n-2} r_{h}^{n-2}.
\end{align}
We find that these quasilocal thermodynamic variables agree with the $n$-dimensional quasilocal Smarr relation (\ref{eq:QLSmarr}),
\begin{equation*}
(n-2)E = (n-2) T_R S + (n-3) \Phi_R Q + (n-2) PA,
\end{equation*}
where
\begin{equation}
A = \omega_{n-2}R^{n-2} 
\end{equation} 
and the variation of the energy $E$ with respect to $S$, $A$, and $Q$ exactly takes a form of the first law of thermodynamics as follows
\begin{align}
dE = T_{\textrm{R}} dS + P dA + \Phi_{R} Q.
\end{align}
Please note that quasilocal Smarr relation also can be reproduced by equating the Euclidean action value with the definition of a thermodynamic potential. Namely, since the Euclidean on-shell action value is calculated as 
\begin{align}
F_{R} =& \frac{1}{(n-2)} E - \frac{1}{(n-2)}\Phi_{R} Q + PA \nonumber\\
\label{eq:FRnD}
\end{align}
and the thermodynamic potential is defined by (\ref{eq:freeEnD}), equating them yields the $n$-dimensional quasilocal Smarr relation (\ref{eq:QLSmarr}). 

\section{Einstein-Maxwell-Dilaton Theory}
Let us now consider Einstein-Maxwell theory with a dilaton coupling
\begin{equation}
\mathcal{L} = \sqrt{-g} \bigg( \frac{1}{\kappa^{2}}R - \frac{1}{2}(\partial \phi)^{2} - \frac{1}{4}e^{ \kappa a \phi} {\mathcal{F}}^{2} \bigg).
\end{equation}
The equations of motion are following
\begin{align}
&R_{\mu \nu} - \frac{1}{2} g_{\mu \nu} R = \frac{\kappa^{2}}{2} T_{\mu \nu}, \\&T_{\mu \nu} =\bigg(\partial_{\mu} \phi \partial_{\nu} \phi - \frac{1}{2} g_{\mu \nu} (\partial \phi)^{2} \bigg) +  e^{\kappa a \phi} \bigg({\mathcal{F}}_{\mu \gamma} {\mathcal{F}}_{\nu}^{\; \; \gamma} - \frac{1}{4} g_{\mu \nu} {\mathcal{F}}^{2} \bigg), \\
&\Box \phi = \frac{1}{4} \kappa a e^{ \kappa a \phi} {\mathcal{F}}^{2}, \qquad \nabla_{\mu} e^{\kappa a \phi} {\mathcal{F}}^{\mu \nu} = 0.
\label{eq:EMDeom}
\end{align}
The solutions of the equations of motion (\ref{eq:EMDeom}) are studied in \cite{Cvetic:1996gq,Duff:1996hp}. Here we firstly investigate the purely electrically or magnetically charged black hole solution with dilaton and then the dyonic black hole with dilaton.

\subsection{Purely electrically or magnetically charged solutions with dilaton}
For either purely electrically or magnetically charged cases with dynamic dilaton, black hole solutions are studied in \cite{Duff:1996hp,Cvetic:1996gq,Geng:2018jck}. The metric solution for both cases is written as
 \begin{align}
ds^{2} = - H^{-\frac{2}{a^{2}+1}} f dt^{2} + H^{\frac{2}{a^{2}+1}} \left(\frac{dr^{2}}{f} + r^{2} d \Omega_{2}^{2} \right), \; \; \; H = 1 + \frac{\mu s^{2}}{r}, \; \; \; f = 1 - \frac{\mu}{r}
\label{eq:RNDbhanstz}
\end{align}
and the solutions for matter fields are respectively given by
\begin{align}
{\mathcal{F}} &= \frac{2}{\kappa}\frac{q}{r^{2}} H^{-2} dt \wedge dr, \qquad \phi = \frac{1}{\kappa} \frac{2a}{a^{2}+1}\log H, \qquad q = \frac{\mu}{\sqrt{a^{2} + 1}} c s
\label{eq:solDRNbh} \\
{\mathcal{F}} &= \frac{2}{\kappa} p \Omega_{2}, \qquad \phi = - \frac{1}{\kappa} \frac{2a}{a^{2}+1}\log H, \qquad p = \frac{\mu}{\sqrt{a^{2} + 1}} c s
\end{align}
where $c= \cosh \delta$ and $s=\sinh \delta$. When $a=0$ case, this solution becomes the RN black hole solution by using a simple coordinate transformation which is studied in section 3 and so we exclude $a=0$ case here. 
The electric or magnetic charge is defined as
\begin{equation}
Q_{\textrm{e}}= \int e^{\kappa a \phi} *{\mathcal{F}} \; , \qquad Q_{\textrm{m}} = \int {\mathcal{F}}.
\end{equation}
Employing the MM-counterterm in this coordinate, it takes a form of 
\begin{equation}
{\hat{K}}_{ab} = \left(r+ \frac{\mu s^{2}}{1+a^{2}} \right) \sigma_{ab}.
\label{eq:Khat2}
\end{equation}
From the renormalized gravity action, the quasilocal energy($E$) and a quasilocal surface pressure($P$) are computed
\begin{align}
E =& \frac{16 \pi}{\kappa^{2}} \bigg[-\sqrt{1-\frac{\mu}{R}} \left(R + \frac{a^{2} s^{2} \mu}{a^{2}+1} \right) \left(1+\frac{s^{2} \mu}{R} \right)^{-\frac{a^{2}}{a^{2}+1}} + \left(R +\frac{s^{2} \mu}{a^{2}+1} \right) \bigg], \\
P =& \frac{1}{\kappa^{2} R^{2}} \bigg[- \frac{2R - \mu}{\sqrt{1-\frac{\mu}{R}}} \left(1 + \frac{s^{2}\mu}{R} \right)^{-\frac{1}{a^{2}+1}} +2 \left(R + \frac{s^{2} \mu}{a^{2}+1} \right) \left(1 + \frac{s^{2} \mu}{R} \right)^{-\frac{2}{a^{2}+1}} \bigg],
\end{align}
and a quasilocal surface($A$) is easily read
\begin{align}
A =& 4 \pi R^{2} \left(1 + \frac{s^{2} \mu}{R} \right)^{\frac{2}{a^{2}+1}}. 
\end{align}
When taking $R \rightarrow \infty$, the energy and the pressure become
\begin{equation}
E = \frac{1}{2} \mu \left(1 + \frac{2}{1+a^{2}} s^{2} \right), \qquad P = 0
\end{equation}
which agree with one in {\cite{Geng:2018jck}}. The gauge field potential and charge in a quasilocal frame are calculated as
\begin{align}
&\Phi_{\textrm{R,e}} = \pm \frac{1}{N(R)} \frac{2}{\kappa} \left(\frac{q}{c^{2} \mu} - \frac{q}{R + s^{2} \mu} \right),\qquad Q_{\textrm{e}} = \pm \frac{8 \pi}{\kappa} q \\
&\Phi_{\textrm{R,m}} = \pm \frac{1}{N(R)} \frac{2}{\kappa} \left(\frac{p}{c^{2} \mu} - \frac{p}{R + s^{2} \mu} \right),\qquad Q_{\textrm{m}} = \pm \frac{8 \pi}{\kappa} p.
\end{align}
Here the magnetic potential is obtained by the dual electric field via the electro-magnetic duality by using (\ref{eq:emdual}) and (\ref{eq:QLMP}). Tolman temperature is  
\begin{align}
T_{\textrm{R}} = \frac{1}{N(R)} \frac{1}{4 \pi \mu} c^{-\frac{4}{a^{2}+1}}. 
\end{align}
The Euclidean action is expressed as
\begin{align}
I_{E} =& -\int_{\mathcal{M}} \sqrt{g} \bigg( \frac{1}{\kappa^{2}}  R - \frac{1}{2} (\partial{\phi})^{2}- \frac{1}{4} e^{\kappa a \phi} {\mathcal{F}}^{2} \bigg) - \frac{2}{\kappa^{2}} \int_{\partial \mathcal{M}} \sqrt{h}(K-\hat{K}).
\end{align}
For an electrically charged case, this yields a thermodynamic potential as follows
\begin{align}
F_{\textrm{R}} \equiv& I_{E} T_{\textrm{R}} =\frac{1}{2} E - \frac{1}{2} \Phi_{\textrm{R},e} Q_{e} + PA \\
\equiv& E - T_{\textrm{R}}S - \Phi_{\textrm{R},e}Q_{e}.
\end{align}
Then we can compute the entropy 
\begin{align}
S = - \frac{F_{R} - (E-\Phi_{\textrm{R},e} Q_{e})}{T_{R}}  = \frac{16 \pi^{2} \mu^{2}}{\kappa^{2}} c^{\frac{4}{a^{2}+1}}
\label{eq:pureEMetrp}
\end{align}
which agrees with the Bekenstein-Hawking entropy. 

For a magnetically charged case, the Euclidean on-shell action yields the quasilocal free energy
\begin{equation}
F_{R} \equiv I_{E} T_{\textrm{R}} = E - T_{R} S
\end{equation}
and the entropy is obtained as
\begin{align}
S = - \frac{F_{R} - E}{T_{R}}  = \frac{16 \pi^{2} \mu^{2}}{\kappa^{2}} c^{\frac{4}{a^{2}+1}}
\end{align}
which is the same as (\ref{eq:pureEMetrp}). 

Since there is an electro-magnetic duality in four dimensional spacetime, a magnetic charge has the same length dimension as an electric charge and contributes to the Smarr relation in the same way of the electrically charged case. Thus above quasilocal thermodynamic variables satisfy the following Smarr relation that is obtained in (\ref{eq:quasiSmarr})
\begin{equation}
E = 2 T_R S + 2 PA + \Phi_{\textrm{R,e}} Q_{\textrm{e}} \; ( \textrm{or} \; \; \Phi_{\textrm{R,m}} Q_{\textrm{m}} )
\end{equation}
which are for the electrically (or magnetically) charged case.  

If varying the quasilocal energy with respect to $S, Q_{e}$ (or $Q_{m}$), and $A$, the following form
\begin{equation}
dE = T_{\textrm{R}} dS + P dA + \Phi_{\textrm{R,e}} dQ_{\textrm{e}} \; ( \textrm{or} \; \; \Phi_{\textrm{R,m}} dQ_{\textrm{m}} ). 
\label{eq:flrtn2}
\end{equation}
is only satisfied up to $1/R$ order in the large $R$ expansion for the variation of $S$ and $Q$'s and up to $1/R^2$ order for the variation of $R$. Thus the first law (\ref{eq:flrtn2}) is not satisfied for any $R$ and this is because of the non-trivial contribution of the Dilaton field that is distributed in the bulk. The dilaton field contribution can be taken into account  as follows
\begin{align}
&\phi_{\textrm{R}}= \frac{1}{\kappa} \frac{2a}{a^{2}+1}\log H(R), \\
&\Pi_{\textrm{R}} =\frac{4 \pi a (-\sqrt{4 \left(a^2+1\right) q^2+\mu ^2} +\mu)}{(1+a^{2})\kappa} \bigg[ \bigg(1+ \frac{\sqrt{4 \left(a^2+1\right) q^2+\mu ^2} - \mu}{2R} \bigg)^{-\frac{a^{2}}{1+a^{2}}} \sqrt{1-\frac{\mu}{R}} - 1 \bigg]
\end{align}
where $\Pi_{\textrm{R}}$ is taken to be zero at infinity. Adding these to the previous one, the following first law is satisfied at any $R$
\begin{align}
&dE = T_{\textrm{R}} dS + P dA + \Phi_{\textrm{R,e}} dQ_{\textrm{e}} (+ \Phi_{\textrm{R,m}} dQ_{\textrm{m}}) + \Pi_{\textrm{R}} d\phi_{\textrm{R}}. 
\label{eq:flrtn3}
\end{align}
When taking $R \rightarrow \infty$, the Smarr relation and the first law are simply approximated to
\begin{align}
&M = 2 T_{\textrm{H}} S + \Phi_{\textrm{e}} Q_{\textrm{e}} \; ( \textrm{or} \; \; \Phi_{\textrm{m}} Q_{\textrm{m}} ), \\
&dM = T_{\textrm{H}} dS + \Phi_{\textrm{e}} dQ_{\textrm{e}}  \;( \textrm{or} \; \; \Phi_{\textrm{m}} dQ_{\textrm{m}} ). 
\end{align}

\subsection{Dyonic solutions with dilaton}

The equations of motion (\ref{eq:EMDeom}) admit the dyonic black hole either for constant dilaton $\phi_{0}$ and running dilaton $\phi$ with specific values of coupling constant, which is $a=0,1,$ and $\sqrt{3}$. The case of $a=0$ is nothing but the dyonic RN black hole which is studied in section 3 and so we do not repeat here. For the rest of the cases, we examine their thermodynamic properties in a quasilocal frame. 

\subsubsection{Constant dilaton $\phi_{0}$}
For a constant dilaton $\phi_{0}$, a dyonic black hole is given by
\begin{align}
&ds^{2} = - f(r) dt^{2} + \frac{dr^{2}}{f(r)} + r^{2} d \Omega_{2}^{2},\qquad f(r) = 1 - \frac{2 M}{r} + \frac{qp}{2 r^{2}}, \\
&F = \frac{1}{\kappa} \bigg(e^{-a \phi_{0}}\frac{q}{r^{2}} dt \wedge dr + p \Omega_{2} \bigg), \\
&\phi = \phi_{0}, \qquad q = \sqrt{2} q_{0}e^{\frac{a}{2}\phi_{0}}, \qquad p = \sqrt{2} q_{0}e^{-\frac{a}{2}\phi_{0}}.
\end{align}
The MM-counterterm is taken for the renormalized gravity action and then the quasilocal variables are computed as
\begin{align}
&E = \frac{16 \pi R}{\kappa^{2}} \bigg(1-N(R) \bigg),\\
&P = \frac{2}{\kappa^{2} R} \bigg[1 -\frac{1}{N(R)} \left(1-\frac{M}{R} \right) \bigg], \qquad A = 4 \pi R^{2}, \\
&T_{\textrm{R}} = \frac{1}{N(R)} \frac{1}{4 \pi } \bigg(\frac{2M}{r_{h}^{2}}-\frac{pq}{r_{h}^{3}} \bigg), \qquad S = \frac{16 \pi^{2} r_{h}^{2}}{\kappa^{2}}, \\
&\Phi_{\textrm{R,e}} =  \frac{1}{N(R)} \frac{1}{\kappa} e^{- a \phi_{0}} \bigg(\frac{q}{r_{h}} -\frac{q}{R} \bigg), \qquad Q_{\textrm{e}} = \frac{4 \pi}{\kappa} q,\\
&\Phi_{\textrm{R,m}} =  \frac{1}{N(R)} \frac{1}{\kappa} e^{a \phi_{0}} \bigg(\frac{p}{r_{h}} -\frac{p}{R} \bigg), \qquad Q_{\textrm{m}} = \frac{4 \pi}{\kappa} p
\end{align}
where the magnetic potential $\Phi_{\textrm{R,m}}$ is obtained via the electro-magnetic duality (\ref{eq:emdual}) and (\ref{eq:QLMP}). If taking $R \rightarrow \infty$, the quasilocal energy and the pressure approach
\begin{equation}
E = \frac{16 \pi M}{\kappa^{2}}, \qquad P = 0.
\end{equation}
The Euclidean on-shell action value yields the quasilocal thermodynamic potential as follows
\begin{align}
F_{\textrm{R}} \equiv& I_{E} T_{\textrm{R}} =\frac{1}{2} E - \frac{1}{2} \Phi_{\textrm{R,e}} Q_{\textrm{e}} + \frac{1}{2} \Phi_{\textrm{R,m}} Q_{\textrm{m}} + PA \\
\equiv& E - T_{\textrm{R}}S - \Phi_{\textrm{R,e}}Q_{\textrm{e}},
\end{align}
and the entropy ($S$) in a quasilocal frame is obtained 
\begin{equation}
S = \frac{F_R - (E - \Phi_{\textrm{R,e}}Q_{\textrm{e}})}{T_R}
\end{equation}
which agrees with the Bekenstein-Hawking entropy. These quasilocal quantities satisfy the Smarr relation and the first law as follows
\begin{align}
&E = 2 T_{\textrm{R}} S + 2 PA + \Phi_{\textrm{R,e}} Q_{\textrm{e}} + \Phi_{\textrm{R,m}} Q_{\textrm{m}} ,\\
&dE = T_{\textrm{R}} d S + PdA +  \Phi_{\textrm{R,e}} d Q_{\textrm{e}} + \Phi_{\textrm{R,m}} d Q_{\textrm{m}}.
\end{align}

\subsubsection{Dilaton coupling $a=1$}
The dyonic black hole solution with the dilaton coupling $a=1$ case takes a form 
\begin{align}
&ds^{2} = - (H_{1}H_{2})^{-1} f dt^{2} + (H_{1}H_{2}) \bigg(\frac{dr^{2}}{f} + r^{2} d \Omega^{2}_{2} \bigg).\\
&F = \frac{1}{\kappa} \bigg(\frac{q}{r^{2}}H_{1}^{-2} dt \wedge p \Omega_{2} \bigg), \qquad \phi = \frac{1}{\kappa} \log \frac{H_{1}}{H_{2}},\\
&f = 1 - \frac{\mu}{r}, \; \; \; H_{i} = 1 + \frac{\mu s_{i}^{2}}{r}, \\
& q = \sqrt{2}\mu s_{1}c_{1}, \; \; \; p = \sqrt{2}\mu s_{2}c_{2} \label{eq:dyonica1}
\end{align}
where $c_{i} = \cosh \delta_{i}$ and $s_{i} = \sinh \delta_{i}$. The renormalized gravity action can be constructed by using the MM-counterterm (\ref{eq:Khat2}). The quasilocal variables are computed as 
\begin{align}
&E = \frac{8 \pi R}{\kappa^{2}} \left(2 + \frac{\mu s_{1}^{2}}{R} + \frac{\mu s_{2}^{2}}{R} \right) \bigg(1- N(R) \bigg), \\
&P = \frac{R}{\kappa^{2} (R + \mu s_{1}^{2})(R + \mu s_{2}^{2})} \bigg[2 + \frac{\mu s_{1}^{2}}{R} + \frac{\mu s_{2}^{2}}{R} - \frac{1}{N(R)} \bigg(2- \frac{\mu}{R} \bigg) \bigg], \\
&A = 4 \pi (R + \mu s_{1}^{2})(R + \mu s_{2}^{2}), \\
&T_{\textrm{R}} = \frac{1}{N(R)}\frac{1}{4 \pi c_{1}^{2}c_{2}^{2} \mu}, \qquad S = \frac{16 \pi^{2} c_{1}^{2} c_{2}^{2} \mu^{2}}{\kappa^{2}}, \\
&\Phi_{\textrm{R},e} = \frac{1}{N(R)} \frac{1}{\kappa} \bigg(\frac{q}{\mu c_{1}^{2}} - \frac{q}{R + \mu s_{1}^{2}} \bigg), \qquad Q_{e} = \frac{4 \pi}{\kappa} q, \\
&\Phi_{\textrm{R},m} =  \frac{1}{N(R)} \frac{1}{\kappa} \bigg(\frac{p}{\mu c_{2}^{2}} - \frac{p}{R + \mu s_{2}^{2}} \bigg), \qquad Q_{m} = \frac{4 \pi}{\kappa} p.
\end{align}
In which $s_1, c_1, s_2$ and $c_2$ can be replaced in terms of $\mu, q$ and $p$ by using (\ref{eq:dyonica1}) 
\begin{equation}
c_{1}^{2} -1 =  s_{1}^{2} = -\frac{1}{2} + \frac{\sqrt{2q^{2} + \mu^{2}}}{2 \mu}, \qquad c_{2}^{2} -1= s_{2}^{2} = -\frac{1}{2} + \frac{\sqrt{2p^{2} + \mu^{2}}}{2 \mu}.
\end{equation}
The Euclidean on-shell action value yields the quasilocal thermodynamic potential as follows
\begin{align}
F_{\textrm{R}} \equiv& I_{E} T_{\textrm{R}} =\frac{1}{2} E - \frac{1}{2} \Phi_{\textrm{R,e}} Q_{\textrm{e}} + \frac{1}{2} \Phi_{\textrm{R,m}} Q_{\textrm{m}} + PA \\
\equiv& E - T_{\textrm{R}}S - \Phi_{\textrm{R,e}}Q_{\textrm{e}},
\end{align}
and then the entropy ($S$) in a quasilocal frame can be found
\begin{equation}
S = \frac{F_R - (E - \Phi_{\textrm{R,e}}Q_{\textrm{e}})}{R}
\end{equation}
which is nothing but the Bekenstein-Hawking entropy. These variables satisfy the following Smarr relation
\begin{align}
&E = 2 T_{\textrm{R}} S + 2 PA + \Phi_{\textrm{R,e}} Q_{\textrm{e}} + \Phi_{\textrm{R,m}} Q_{\textrm{m}}. 
\end{align}
If varying the quasilocal energy with respect to $S, Q_{\textrm{e}}, Q_{\textrm{m}}$ and $R$, the following form 
\begin{align}
d E = T_{\textrm{R}} dS + P dA + \Phi_{\textrm{R,e}} dQ_{\textrm{e}} + \Phi_{\textrm{R,m}} dQ_{\textrm{m}} 
\end{align}
is satisfied up to $1/R$ order for the large $R$ expansion for the variation with respect to $S$ and $Q$'s and up to $1/R^2$ order for the variation with respect to $A$ due to the dilaton contribution in the bulk. The dilaton field and its conjugate potential take forms of
\begin{align}
&\phi_{\textrm{R}} = \frac{1}{\kappa} \log \frac{H_{1}(R)}{H_{2}(R)}, \\ 
&\Pi_{\textrm{R}} = \frac{2 \pi (x_{2}-y_{2})}{\kappa} \bigg[ -\frac{2 R \sqrt{1-\frac{\mu}{R}}}{\sqrt{(x_{2}+ 2R - \mu)}\sqrt{(y_{2}+ 2R - \mu)}} +1 \bigg]
\end{align}
where $x_{2} = \sqrt{2q^{2} + \mu^{2}}$ and $y_{2} = \sqrt{2p^{2} + \mu^{2}}$ and $\Pi_{\textrm{R}}$ is taken to be zero at infinity. If adding this dilaton contribution, the following the first law is satisfied for all orders of $R$
\begin{align}
d E = T_{\textrm{R}} dS + P dA + \Phi_{\textrm{R,e}} dQ_{\textrm{e}} + \Phi_{\textrm{R,m}} dQ_{\textrm{m}} + \Pi_{\textrm{R}} d \phi_{\textrm{R}}.
\end{align}

As $R$ approaches the infinity, the quasilocal energy, pressure, and the Tolman temperature become
\begin{equation}
E = \frac{8 \pi}{\kappa^{2}} \mu(1+ s_{1}^{2} + s_{2}^{2}) = M, \qquad P = 0 \qquad T_R = T_H
\end{equation}
which agrees with ones in \cite{Geng:2018jck} and the following Smarr and the first law are satisfied
\begin{align}
&M = 2 T_H S + \Phi_{\textrm{e}} Q_{\textrm{e}} + \Phi_{\textrm{m}} Q_{\textrm{m}},\\
&dM = T_H dS + + \Phi_{\textrm{e}} dQ_{\textrm{e}} + \Phi_{\textrm{m}} dQ_{\textrm{m}} . 
\end{align}

\subsubsection{Dilaton coupling $a=\sqrt{3}$}
The dyonic black hole solution with the dilaton coupling $a=\sqrt{3}$ case takes a form 
\begin{align}
&ds^{2} = - (H_{1}H_{2})^{-\frac{1}{2}} f dt^{2} + (H_{1}H_{2})^{\frac{1}{2}} \bigg(\frac{dr^{2}}{f} + r^{2} d \Omega^{2}_{2} \bigg).\\
&F = \frac{1}{\kappa} \bigg(\frac{q}{r^{2}}H_{1}^{-2} H_{2} dt \wedge p \Omega_{2} \bigg), \qquad \phi = \frac{\sqrt{3}}{2 \kappa} \log \frac{H_{1}}{H_{2}},\\
&f = 1 - \frac{\mu}{r}, \qquad H_{i} = 1 + \frac{\mu s_{i}^{2}}{r} + \frac{\mu^{2} c_{i}^{2} s_{1}^{2} s_{2}^{2}}{2(c_{1}^{2}+c_{2}^{2})r^{2}}, \\
&q =\mu s_{1} c_{1} \sqrt{\frac{1+c_{1}^{2}}{c_{1}^{2} + c_{2}^{2}}}, \qquad p = \mu s_{2} c_{2} \sqrt{\frac{1+c_{2}^{2}}{c_{1}^{2} + c_{2}^{2}}}.
\end{align}
where $c_{i} = \cosh \delta_{i}$ and $s_{i} = \sinh \delta_{i}$. The same as before, the MM-counterterm (\ref{eq:Khat2}) can be used to build the renormalized gravity action. Then we can compute all thermodynamic variables as follows
\begin{align}
&E = \frac{4 \pi}{\kappa^{2}} \bigg[4R + (s_{1}^{2}+s_{2}^{2})\mu -  \sqrt{1-\frac{\mu}{R}} \frac{H_{1}(R)(2 R + s_{2}^{2} \mu) + H_{2}(R) (2 R + s_{1}^{2} \mu)}{(H_{1}(R) H_{2}(R) )^{\frac{3}{4}} } \bigg],\\
&P = \frac{1}{2 \kappa^{2}} \frac{1}{(H_{1}(R) H_{2} (R))^{\frac{1}{2}} R^{2}} \bigg[ 4R + (s_{1}^{2}+s_{2}^{2}) \mu - \frac{1}{N(R)} \bigg(R-\frac{\mu}{2} \bigg) \bigg], \\
&A = 4 \pi R^{2} (H_{1}(R) H_{2}(R))^{\frac{1}{2}}, \\
&T_{\textrm{R}} = \frac{1}{N(R)} \frac{1}{4 \pi \mu} \frac{1}{c_{1} c_{2}} H_{0}^{-1}, \qquad S = \frac{8 \pi^{2} \mu^{2}}{\kappa^{2}} c_{1} c_{2} H_{0}, \\
&\Phi_{\textrm{R,e}} = \frac{q}{4 \kappa} \bigg(\frac{2+s_{2}^{2}}{H_{0} c_{1}^{2} \mu} - \frac{2R + s_{2}^{2} \mu}{R^{2} H_{1}(R)} \bigg), \qquad Q_{\textrm{e}} = \frac{4 \pi}{\kappa} q, \\
&\Phi_{\textrm{R,e}} = \frac{p}{4 \kappa} \bigg(\frac{2+s_{1}^{2}}{H_{0} c_{2}^{2} \mu} - \frac{2R + s_{1}^{2} \mu}{R^{2} H_{2}(R)} \bigg) , \qquad Q_{\textrm{m}} = \frac{4 \pi}{\kappa} p
\end{align}
where $H_{0} = 1+ \frac{s_{1}^{2} s_{2}^{2}}{2(c_{1}^{2}+ c_{2}^{2})}$. The Euclidean on-shell action value yields the quasilocal thermodynamic potential as follows
\begin{align}
F_{\textrm{R}} \equiv& I_{E} T_{\textrm{R}} =\frac{1}{2} E - \frac{1}{2} \Phi_{\textrm{R,e}} Q_{\textrm{e}} + \frac{1}{2} \Phi_{\textrm{R,m}} Q_{\textrm{m}} + PA \\
\equiv& E - T_{\textrm{R}}S - \Phi_{\textrm{R,e}}Q_{\textrm{e}}
\end{align}
and then entropy ($S$) in a quasilocal frame is calculated 
\begin{equation}
S = \frac{F_R - (E-\Phi_{\textrm{R,e}}Q_{\textrm{e}})}{T_R}
\end{equation}
which turns to agree with the Bekenstein-Hawking entropy. As expected, these quasilocal variables satisfy the following Smarr relation 
\begin{equation}
E = 2 T_{\textrm{R}} S + 2 PA + \Phi_{\textrm{R,e}} Q_{\textrm{e}} + \Phi_{\textrm{R,m}} Q_{\textrm{m}}. 
\label{eq:SMasqrt3}
\end{equation}
To make a variation of $\delta_1 = \delta_1(\mu, q, p)$ and $\delta_2 = \delta_2(\mu, q, p)$ with respect to $\mu, q ,p$ and $R$, it is useful to use 
\begin{align}
&\partial_{\mu} c_1^2 =-\frac{c_1^2 \left(c_1^4-1\right) \left(3 c_2^4-1\right)}{\left(\left(3 c_2^4-1\right) c_1^4+c_2^2 c_1^2-c_2^4\right) \mu }, \qquad \partial_{\mu} c_2^2 = -\frac{\left(3 c_1^4-1\right) \left(c_2^4-1\right) \text{c2}^2}{\left(\left(3 c_2^4-1\right) c_1^4+c_2^2 c_1^2-c_2^4\right) \mu }, \\
&\partial_{q} c_1^2 =\frac{c_1^2 \left(c_1^4-1\right) \left(\left(3 c_2^4-1\right) \text{c1}^2+2 c_2^6\right)}{\left(\left(3 c_2^4-1\right) c_1^6+3 c_2^6 c_1^4-c_2^6\right) q}, \qquad \partial_{q} c_2^2 = \frac{c_1^2 \left(c_1^4-1\right) \left(c_2^4-1\right) \text{c2}^2}{\left(\left(3 c_2^4-1\right) c_1^6+3 c_2^6 c_1^4-c_2^6\right) q}, \\
&\partial_{p} c_1^2 =\frac{c_1^2 \left(c_1^4-1\right) c_2^2 \left(c_2^4-1\right)}{\left(\left(3 c_2^4-1\right) c_1^6+3 c_2^6 c_1^4-c_2^6\right) p}, \qquad \partial_{p} c_2^2 =\frac{c_2^2 \left(c_2^4-1\right) \left(2 c_1^6+3 c_2^2 c_1^4-c_2^2\right)}{\left(\left(3 c_2^4-1\right) c_1^6+3 c_2^6 c_1^4-c_2^6\right) p}
\end{align}
where $\partial s_1^2 = \partial c_1^2$ and $\partial s_2^2 = \partial c_2^2$. As we learned about the first law from the previous examples having dynamic dilaton field, here we have the same consequence. The following form of the first law 
\begin{equation}
dE =  T_{\textrm{R}} d S + P dA + \Phi_{\textrm{R,e}} dQ_{\textrm{e}} + \Phi_{\textrm{R,m}} d Q_{\textrm{m}} 
\end{equation}
is only valid up to the order of $1/R$ for the variation on $\mu, q,$ and $p$ and up to the order of $1/R^2$ for the variation on $R$ when expanding on large $R$. Then as we did in the previous examples we plug the dilaton field and its conjugate potential, which is obtained from (\ref{eq:PiR})
\begin{align}
&\phi_{\textrm{R}} = \frac{\sqrt{3}}{2 \kappa} \log \frac{H_{1}(R)}{H_{2}(R)}, \\
&\Pi_{\textrm{R}} =  \frac{2\sqrt{3}\pi}{\kappa} \bigg[ (s_{1}^{2} - s_{2}^{2}) \mu + (H_{1}(R) H_{2}(R))^{1/4} \sqrt{1-\frac{\mu}{R}} \bigg(\frac{2 R+ \mu  s_1^2}{H_1(R)}  -  \frac{2 R+ \mu  s_2^2}{H_2(R)}\bigg)   \bigg] \label{eq:PiRasq3}
\end{align}
into the first law as follows
\begin{equation}
dE =  T_{\textrm{R}} d S + P dA + \Phi_{\textrm{R,e}} dQ_{\textrm{e}} + \Phi_{\textrm{R,m}} d Q_{\textrm{m}} + \Pi_{\textrm{R}} d\phi_{\textrm{R}}. \label{eq:flasq3}
\end{equation}
However, this dilaton potential only improves the variation of $R$ up to the order of $1/R^3$ in this case. They do not capture all the contributions of dilaton field in the bulk but has the remnant as follows
\begin{align}
&d_{\mu}E -(T_{\textrm{R}} d_{\mu} S + P d_{\mu}A + \Phi_{\textrm{R,e}} d_{\mu} Q_{\textrm{e}} + \Phi_{\textrm{R,m}} d_{\mu} Q_{\textrm{m}} + \Pi_{\textrm{R}} d_{\mu} \phi_{\textrm{R}}) \nonumber\\
&\qquad \qquad \sim \frac{\pi  \mu ^2 s_{1}^2 s_{2}^2 \left( c_{1}^2 + 1 \right) \left(c_{2}^2 + 1 \right) \left(c_{1}^2-c_{2}^2\right)^2}{\kappa ^2 \left(c_1^2+c_2^2\right) \left(-3 c_2^4 c_1^4+c_1^4-c_2^2 c_1^2+c_2^4\right) R^{2}} + \cdots \label{eq:dmuEasqrt3}\\
&d_{q}E -(T_{\textrm{R}} d_{q} S + P d_{q}A + \Phi_{\textrm{R,e}} d_{q} Q_{\textrm{e}} + \Phi_{\textrm{R,m}} d_{q} Q_{\textrm{m}} + \Pi_{\textrm{R}} d_{q} \phi_{\textrm{R}})  \nonumber\\
&\qquad \qquad \sim \frac{\pi  \mu^2 c_{1} s_{1} s_{2}^2 \left(c_{2}^2+1\right)  \left(c_{1}^2-c_{2}^2\right) \left(c_{1}^2 - c_{2}^2 - 2 c_{1}^2 c_{2}^4 \right)}{\kappa^{2}\left(c_{1}^2+c_{2}^2\right) \left(c_{1}^4 \left(3 c_{2}^4-1\right)+c_{1}^2 c_{2}^2-c_{2}^4\right) R^{2}} \sqrt{\frac{1+c_{1}^{2}}{c_{1}^{2} +c_{2}^{2}}} + \cdots \label{eq:dqEasqrt3}\\\
&d_{p}E -(T_{\textrm{R}} d_{p} S + P d_{p}A + \Phi_{\textrm{R,e}} d_{p} Q_{\textrm{e}} + \Phi_{\textrm{R,m}} d_{p} Q_{\textrm{m}} + \Pi_{\textrm{R}} d_{p} \phi_{\textrm{R}}) \nonumber\\
& \qquad \qquad \sim \frac{\pi \mu^2  c_{2} s_{1}^2 s_{2}\left(c_{1}^2+1\right) \left(c_{1}^2-c_{2}^2\right) \left(c_{1}^2-c_{2}^2 + 2 c_{1}^4 c_{2}^2\right)}{\kappa ^2  \left(c_{1}^2+c_{2}^2\right) \left(c_{1}^4 \left(3 c_{2}^4-1\right)+c_{1}^2 c_{2}^2-c_{2}^4\right)R^2} \sqrt{\frac{1+c_{2}^{2}}{c_{1}^{2} +c_{2}^{2}}} + \cdots  \label{eq:dpEasqrt3}\ \\
&d_{R}E -(T_{\textrm{R}} d_{R} S + P d_{R}A + \Phi_{\textrm{R,e}} d_{R} Q_{\textrm{e}} + \Phi_{\textrm{R,m}} d_{R} Q_{\textrm{m}} + \Pi_{\textrm{R}} d_{R} \phi_{\textrm{R}})  \nonumber\\
&\qquad \qquad \sim -\frac{\pi  \mu ^4 s_{1}^2 s_{2}^2   \left(c_{1}^2+1\right) \left(c_{2}^2+1\right)\left(c_{1}^2-c_{2}^2\right)^2}{\kappa ^2 \left(c_{1}^2+c_{2}^2\right)^2 R^4 } + \cdots \label{eq:dREasqrt3}\
\end{align}
where the dots indicate the higher order of $1/R$. Then we might give up the definition of the dilaton potential (\ref{eq:PiR}) and see if another definition of $\Pi_R$ can make the the first law satisfied. So we imposed the first law (\ref{eq:flasq3}) to hold and then found $\Pi_R$ from each variation of the parameters. However the values of $\Pi_R$ obtained from the first law were not consistent. Another possibility is to consider an unknown contribution such as $XdY$, but it also seems not to be the case. This is because $XdY$ term should be effective from the order of $1/R^2$ seeing in (\ref{eq:dmuEasqrt3}) - (\ref{eq:dpEasqrt3}) but then this will contribute to the order of $1/R^3$ in (\ref{eq:dREasqrt3}), which should not be present. This indicates that we cannot find $XdY$ term to content the first law. Only way to satisfy the first law for all order of $R$ is to set $\delta_1 = \delta_2$, but this is nothing but the condition that turns off the dilaton field. 

Taking $R \rightarrow \infty$, the quasilocal energy and pressure take values 
\begin{equation}
E = \frac{4 \pi}{\kappa^{2}} \mu (c_{1}^{2} + c_{2}^{2} ) = M, \qquad P = 0
\end{equation}
which agree with ones in \cite{Geng:2018jck} and the following Smarr relation and the first law are satisfied
\begin{align}
&M = 2 T_{\textrm{H}} S + \Phi_{\textrm{e}} Q_{\textrm{e}}  + \Phi_{\textrm{m}} Q_{\textrm{m}}, \\
&dM = T_{\textrm{H}} dS + \Phi_{\textrm{e}} dQ_{\textrm{e}}  + \Phi_{\textrm{m}} dQ_{\textrm{m}}. 
\end{align}

\section{Summary and Future work}

We studied thermodynamics in a quasilocal frame for Einstein-Maxwell(-Dilaton) theory in an asymptotically flat spacetime. In order to do so, we extended thermodynamic relations such as thermodynamic potential, Smarr relation and the first law to a quasilocal frame and checked them with various black hole solutions.  

We firstly employed Brown-York quasilocal method with Mann-Marolf counterterm to obtain quasilocal energy and surface pressure from the renormalized action. We also showed the form of electric and magnetic potential in a quasilocal frame. In four dimensional spacetime, they make the same contribution to the Smarr relation due to the electro-magnetic duality. Secondly we defined quasilocal thermodynamic potential from the Euclidean on-shell action by applying Tolman temperature and derived entropy in a quasilocal frame. Here the quasilocal thermodynamic potential does not contain the surface pressure term and the entropy derived from the quasilocal thermodynamic potential agrees with Bekenstein-Hawking entropy. Then we formulated quasilocal Smarr relation from Euler’s theorem and found that the surface pressure term and its conjugate variables participate in the relation. When $r$ goes infinity, the quasilocal energy approaches the conserved charge of the spacetime and the surface pressure vanishes and makes no effect on the thermodynamic relations. Other quasilocal thermodynamic variables also restore the usual thermodynamic form at infinity. 

Under these construction, we investigated the quasilocal thermodynamics for various black hole solutions of  Einstein-Maxwell(-Dilaton) theory and checked if these well describe the quasilocal thermodynamics. Apart from the importance of the surface pressure term, one should also notice the importance of dilaton field in a quasilocal frame. For dyonic black hole solutions having dynamic dilaton, while the quasilocal Smarr relation does not contain the dilaton term, there is additional contribution of the dynamic dilaton field to the first law at finite $R$ but their contribution dies out at infinity. Namely, when we expand the quasilocal first law on large $R$ in the absence of the dilaton field and its conjugate potential term, the quasilocal first law is satisfied up to $1/R$ order for the variation with respect to the entropy and Maxwell charges and up to $1/R^2$ order for the variation with respect to the quasilocal area due to the spread of dilaton field in the bulk. This dilaton contribution can be taken into account by adding the dilaton terms to the first law and then the quasilocal first law is satisfied. However we found the exceptional case which is the dilaton coupling constant $a=\sqrt{3}$. In this case adding the dilaton terms does not capture all the contributions in the bulk and it seems not improved by modifying the dilaton potential or considering a new term such as $XdY$ to the quasilocal first law. The only condition making the first law to consent is to set the electric and magnetic charge equivalent which induces turning off the dilaton field.

This work could extend to find quasilocal Smarr relations for various spacetime using different methods. The Smarr relation is firstly found in \cite{Smarr:1972kt} by applying Euler's theorem, but later the same form is derived from the Komar formula in \cite{Bardeen:1973gs}. In this paper, the quasilocal Smarr relation is obtained by the same way of \cite{Smarr:1972kt}, but it is interesting to see whether the Komar formula can be modified to yield the same quasilocal Smarr relation given here. Also, exploring the quasilocal Smarr relation for black hole spacetimes having a negative cosmological constant is very intriguing. The Smarr relation for AdS black hole and Lifshitz black holes are studied by introducing a pressure term with a thermodynamic volume term in \cite{Kastor:2009wy,Dolan:2011xt,Kubiznak:2014zwa} and \cite{Brenna:2015pqa} respectively. Quasilocal thermodynamics for an AdS black hole is studied in \cite{Creighton:1996st,Brown:1994gs}. However, its extension to the quasilocal Smarr relation is non-trivial and should be pursued. 

\section*{Acknowledgement}

M. Park would like to thank Sang-Heon Yi, Nakwoo Kim, Wonwoo Lee, Jungjai Lee, Jong-Dae Park, Papadimitrious Ioannis, Hong Lu, Mu-In Park, David Kubiz\u{n}\'{a}k and Robert B. Mann for useful discussions. M. Park was supported by a KIAS Individual Grant (PG062001) at Korea Institute for Advanced Study and by Basic Science Research Program through the National Research Foundation of Korea funded by the Ministry of Education (NRF-2016R1D1A1B03933399).

\bibliographystyle{JHEP}

\bibliography{QLSMRR_arXiv_Jan212021}

\end{document}